\documentclass[hidelinks,conference]{IEEEtran}
\IEEEoverridecommandlockouts

\usepackage{cite}
\usepackage{amsmath,amssymb,amsfonts}
\usepackage{graphicx}
\usepackage{textcomp}
\usepackage{xcolor}
\usepackage{orcidlink}
\usepackage[acronym,shortcuts]{glossaries}
\usepackage{bm}
\usepackage{caption}
\usepackage{subfig}
\usepackage[export]{adjustbox}
\usepackage{relsize}
\usepackage{soul}
\usepackage{algorithm,algpseudocode}
\usepackage{relsize} 
\usepackage{outlines}
\usepackage{lipsum}
\usepackage{scalefnt}
\usepackage{physics}
\usepackage{amsthm}
\usepackage[bottom]{footmisc}
\usepackage{stfloats}
\newcommand{\herm}[0]{^{\mathsf{H}}}

\newacronym{AFDM}{AFDM}{affine frequency division multiplexing}
\newacronym{IM}{IM}{index modulation}
\newacronym{RF}{RF}{radio frequency}
\newacronym{ISAC}{ISAC}{integrated sensing and communications}
\newacronym{6G}{6G}{sixth-generation}
\newacronym{DD}{DD}{doubly-dispersive}
\newacronym{CPIM}{CPIM}{chirp-permutation-index modulation}
\newacronym{GAS}{GAS}{Grover adaptive search}
\newacronym{MMSE}{MMSE}{minimum mean-squared-error}
\newacronym{ML}{ML}{maximum likelihood}
\newacronym{SISO}{SISO}{single-input single-output}
\newacronym{AWGN}{AWGN}{additive white Gaussian noise}
\newacronym{CP}{CP}{cyclic prefix}
\newacronym{CPP}{CPP}{chirp-periodic prefix}
\newacronym{OFDM}{OFDM}{orthogonal frequency division multiplexing}
\newacronym{AFT}{AFT}{affine Fourier transform}
\newacronym{DFT}{DFT}{discrete Fourier transform}
\newacronym{IDFT}{IDFT}{inverse discrete Fourier transform}
\newacronym{DAFT}{DAFT}{discrete affine Fourier transform}
\newacronym{IDAFT}{IDAFT}{inverse discrete affine Fourier transform}
\newacronym{BER}{BER}{bit-error-rate}
\newacronym{QUBO}{QUBO}{quadratic unconstrained binary optimization}
\newacronym{LTV}{LTV}{linear time-variant}
\newacronym{TVIRF}{TVIRF}{time-varying impulse response function}
\newacronym{OTFS}{OTFS}{orthogonal time frequency space}
\newacronym{CSI}{CSI}{channel state information}
\newacronym{BPSK}{BPSK}{binary phase shift keying}
\newacronym{QPSK}{QPSK}{quadrature phase shift keying}
\newacronym{5GNR}{5GNR}{5G New Radio}
\newacronym{EbN0}{$E_b/N_0$}{energy-per-bit-to-noise-spectral-density ratio}
\newacronym{SNR}{SNR}{signal-to-noise ratio}

\title{AFDM Chirp-Permutation-Index Modulation \\ with Quantum-Accelerated Codebook Design}

\author{
\IEEEauthorblockN{Hyeon~Seok~Rou\IEEEauthorrefmark{1},
Kein~Yukiyoshi\IEEEauthorrefmark{2},
Taku~Mikuriya\IEEEauthorrefmark{2}, 
Giuseppe~Thadeu~Freitas~de~Abreu\IEEEauthorrefmark{1}, and Naoki Ishikawa\IEEEauthorrefmark{2}} \\[-0.5ex]
\IEEEauthorblockA{\IEEEauthorrefmark{1}\! School of Computer Science and Engineering, Constructor University, Campus Ring 1, 28759, Bremen, Germany 
}
\IEEEauthorblockA{\IEEEauthorrefmark{2}\! Graduate School of Engineering Science, Yokohama National University, 240-8501 Kanagawa, Japan 
}
\vspace{-4.75ex}
}

\begin{document}
\maketitle

\begin{abstract}
We describe a novel \ac{IM} scheme exploiting a unique feature of the recently proposed \ac{AFDM} in 
\ac{DD} channels.
Dubbed \ac{AFDM} \ac{CPIM}, the proposed method encodes additional information via the permutation of the \ac{DAFT} chirp sequence, without any sacrifice of the various beneficial properties of the \ac{AFDM} waveform in \ac{DD} channels.
The effectiveness of the proposed method is validated via simulation results leveraging a novel reduced-complexity \ac{MMSE}-based \ac{ML} detector, highlighting the gains over the classical \ac{AFDM}.
As part of the work two interesting problems related to optimizing \ac{AFDM}-\ac{CPIM} are identified: the optimal codebook design problem, over a discrete solution space of dimension $\binom{N!}{K}$, where $N$ is the number of subcarriers and $K$ is the number of codewords; and the \ac{ML} detection problem whose solution space is of dimension $KM^N$, where $M$ is the constellation size.
In order to alleviate the computational complexity of these problems and enable large-scale variations of \ac{AFDM}-\ac{CPIM}, the two problems are reformulated as a higher-order binary optimization problem and mapped to the well-known quantum \ac{GAS} algorithm for their solution.

\end{abstract}
\glsresetall


\vspace{-0.5ex}

\section{Introduction}
\label{sec:introduction}

\vspace{-0.5ex}

Recently, \ac{AFDM} has emerged as a strong candidate waveform for \ac{6G} wireless systems \cite{Bemani_TWC23,Rou_arxiv24}, as it offers a number of advantages such as robustness to \ac{DD} channels, attractive features for \ac{ISAC}, the ability to minimize self-interference via de-chirping and more \cite{Wang_JCS22}.
On the other hand, \ac{IM} is also an attractive technology for \ac{6G} systems, as it has other desirable (and to some extent complementary) characteristics such as reduction of \ac{RF} chains \cite{Dang_FCN21}, and ability to easily incorporate physical-layer security \cite{Li_Net23}.

In most \ac{IM} schemes, a subset of a larger number of available resources are selected in order to encode information, such that the utilization of available resources is ``sparsified'', which we shall therefore refer to such approaches as sparse \ac{IM} schemes.
As shall be shown later, in Subsection \ref{sec:AFDM-CPIM}, in the case of \ac{AFDM} a very peculiar type of \ac{IM} can also be designed, in which no resource is excluded at any transmission instance\footnotemark.

In particular, we refer to the possibility -- unique to \ac{AFDM} in comparison to alternative waveforms such as \ac{OFDM} and \ac{OTFS} \cite{Wei_WC21} -- that the chips employed in the construction of the \ac{AFDM} chirps be permuted in accordance with the information to be transmitted.
The resulting technique, hereafter refered to as \ac{CPIM}, is one of the main contribution of this article.

As shall be clarified, the consequence of such pseudo-random permutation of \ac{AFDM} chirp chips, is a substantial difference in that the state of the channel between the transmitter and receiver.
To elaborate, it is found that different channels matrices corresponding to a distinct chirp permutations, are very distinguishable, under various definitions of matrix distances.
By designing distinct sets of chip permutations that maximize the distinction among the resulting \ac{AFDM} channels, an optimum \ac{CPIM} codebook design can be achieved.
Denoting the number of subcarriers by $N$, and the number of codewords ($i.e.$ selected permutations) by $K$, it follows that the optimal codebook design problem requires a search over a discrete solution space of dimension $\binom{N!}{K}$. 
In turn, at the receiver, \ac{ML} detection requires that all possible $K$ permutations be tested, each over all possible $N$-tuple of symbols taking from a constellation of size $M$, such that again the search space for \ac{ML} detection is of dimension $KM^N$.

\footnotetext{We emphasize that, of course, sparse \ac{IM} can also be considered in conjunction to the method here proposed. For example, the set of chirp-domain subcarriers employed during a given transmission instance may be taken from a larger subset. Similarly, spatial-domain variations \ac{IM} can also be incorporated. Such approaches are transparent to the proposed method.}

Clearly, such combinatorial search spaces can quickly grow well beyond the feasibility of modern digital computers even for relatively small numbers of $K$, $N$ and $M$, which motivates the second major contribution of the article, namely, a pair of quantum-computing algorithms for both the \ac{ML} detection and codebook design of \ac{AFDM}-\ac{CPIM}.

We remark that although techniques to reduce decoding complexity exist, for instance via sphere-detection \cite{Zheng_TC17}, compressive sensing approaches \cite{An_TVT22}, or vectorized message-passing algorithms \cite{Rou_CAMSAP23}, none of such techniques can actually achieve the same performance of the brute-force \ac{ML} search.
In turn, the codebook design is an even harder problem, related to the well-known max-min dispersion problem \cite{Ravi_op}.
We demonstrate, however, that the codebook design can be first mapped to a binary optimization problem, allowing the query complexity to be reduced to approximately the square root via quantum computing techniques.
Such representative solution is the \ac{GAS} algorithm, which was shown to be efficiently implemented via quantum oracle in the case of objective functions with integer coefficients in \cite{gilliam2021grover}, extended to accommodate real-valued coefficients \cite{norimoto2023quantum}.

\section{System Model}
\label{sec:system_model}


\subsection{Conventional AFDM Signal Model}
\label{sec:AFDM_model}

The \ac{AFDM} waveform \cite{Bemani_TWC23} leverages the \ac{IDAFT} \cite{Healy_LCT15} to modulate a sequence of information symbols unto the chirp-domain subcarriers.

The $N$-point forward \ac{DAFT} matrix $\mathbf{A}\in\mathbb{C}^{N \times N}$ and its inverse $\mathbf{A}^{-1}\in\mathbb{C}^{N \times N}$ are efficiently described by a \ac{DFT} matrix twisted by two chirp sequences, as
\begin{subequations}
\begin{gather}
\mathbf{A} \triangleq \mathbf{\Lambda}_{c_2} \mathbf{F}_N \mathbf{\Lambda}_{c_1} \in \mathbb{C}^{N \times N}, 
\label{eq:DAFT_matrix}\\
\mathbf{A}^{\!-1} \triangleq (\mathbf{\Lambda}_{c_2} \mathbf{F}_N \mathbf{\Lambda}_{c_1})^{-1} =  \mathbf{\Lambda}_{c_1}\herm \mathbf{F}_N\herm  \mathbf{\Lambda}_{c_2}\herm \in \mathbb{C}^{N \times N},
\label{eq:IDAFT_matrix}
\end{gather}
\label{eq:DAFT_matrices}%
\end{subequations}
where $\mathbf{F}_N \times \mathbb{C}^{N \times N}$ is the normalized $N$-point \ac{DFT} matrix, and $\mathbf{\Lambda}_{c_1} \triangleq \mathrm{diag}(\boldsymbol{\lambda}_{c_1}) \in \mathbb{C}^{N \times N}$, $\mathbf{\Lambda}_{c_2} \triangleq \mathrm{diag}(\boldsymbol{\lambda}_{c_2}) \in \mathbb{C}^{N \times N}$ are diagonal chirp matrices whose diagonals are described by the chirp vector $\boldsymbol{\lambda}_{c_i} \triangleq [e^{-j2\pi c_i (0)^2}, \cdots, e^{-j2\pi c_i (N-1)^2}] \in \mathbb{C}^{N \times 1}$ with a central digital frequency of $c_i$.

The first chirp frequency $c_1$ is a critical parameter which is optimally configured to the statistics of the doubly-dispersive channel to obtain the notable orthogonality of the \ac{AFDM} subcarriers over the delay-Doppler domain \cite{Bemani_TWC23,Rou_arxiv24}.
On the other hand, the second chirp frequency $c_2$ is a relatively flexible parameter which does affect the orthogonality of the subcarriers, but changes certain waveform properties such as the ambiguity function for \ac{ISAC} techniques \cite{Zhu_arXiv23}.
The optimal criteria of $c_1$ and $c_2$ will be discussed in the following section.

In light of the above, the \ac{AFDM} modulated transmit signal $\mathbf{s} \in \mathbb{C}^{N \times 1}$ is given by an $N$-point \ac{IDAFT} of the symbol vector $\mathbf{x} \in \mathcal{X}^{N \times 1}$ whose elements are drawn from an $M$-ary complex digital constellation $\mathcal{X} \subset \mathbb{C}$, \textit{i.e.,}
\begin{equation}
\mathbf{s} \triangleq \mathbf{A}^{\!-1} \mathbf{x} = \mathbf{\Lambda}_{c_1}\herm \mathbf{F}_N\herm  \mathbf{\Lambda}_{c_2}\herm \mathbf{x} \in \mathbb{C}^{N \times 1}.
\label{eq:AFDM_modulation}
\end{equation}

\subsection{Received Signal Model over Doubly-Dispersive Channels}
\label{sec:IO_AFDM}

\begin{figure*}[hb!]
\vspace{-2ex}
\hrule
\vspace{1ex}
\begin{equation}
\mathbf{\Phi}_{p} \triangleq \mathrm{diag}\Big(\big[\overbrace{e^{-j2\pi\cdot c_1(N^2 - 2N(\ell_p))}, e^{-j2\pi\cdot c_1(N^2 - 2N(\ell_p - 1))}, \cdots, e^{-j2\pi\cdot c_1(N^2 - 2N(1))}}^{\ell_p \;\text{terms}}, \overbrace{\;\!1\;\!, 1\;\!, \cdots\!\vphantom{e^{(x)}}, 1\;\!, 1}^{N - \ell_p\;\! \text{ones}}\big]\Big) \in \mathbb{C}^{N \times N}\!.
\label{eq:CCP_phase_matrix}
\end{equation}
\begin{equation}
\mathbf{Z} \triangleq \mathrm{diag}\Big(\big[e^{-j2\pi(0)/N}, e^{-j2\pi(1)/N}, \; \cdots, e^{-j2\pi(N-2)/N}, e^{-j2\pi(N-1)/N}\big]\Big) \in \mathbb{C}^{N \times N}.
\label{eq:Z_matrix}
\end{equation}
\end{figure*}

Consider a doubly-dispersive channel between a transmitter and a receiver with $P$ significant resolvable propagation paths, where each $p$-th path induces a unique path delay $\tau_p \in [0, \tau^\mathrm{max}]$ and Doppler shift $\nu_p \in [-\nu^\mathrm{max}, +\nu^\mathrm{max}]$ to the received signal, with the channel statistics described by the delay spread $\tau^\mathrm{max}$ and Doppler spread $\nu^\mathrm{max}$.
The corresponding received signal is obtained by a linear convolution of the transmit signal and the \ac{TVIRF} of the doubly-dispersive channel \cite{Hong_2022, Rou_arxiv24}.

Then, by leveraging a \ac{CP} to the transmit signal, and sampling the signals at a sampling frequency of $f_\mathrm{s} \triangleq \frac{1}{T_\mathrm{s}}$, the received signal can be expressed in terms of a discrete circular convolutional channel given by \vspace{-1.25ex}
\begin{equation}
\mathbf{r} \triangleq \mathbf{H} \mathbf{s}  + \mathbf{w} = \Big( \sum_{p=1}^{P}h_p \!\cdot\! \mathbf{\Phi}_{p} \!\cdot\! \mathbf{Z}^{f_p} \!\cdot\! \mathbf{\Pi}^{\ell_p}\Big)\mathbf{s} + \mathbf{w} \in \mathbb{C}^{N \times 1},
\label{eq:matrix_received_signal}
\vspace{-1.5ex}
\end{equation}
where $\mathbf{r} \in \mathbb{C}^{N \times 1}$, $\mathbf{s} \in \mathbb{C}^{N \times 1}$, and $\mathbf{w} \in \mathbb{C}^{N \times 1}$ are the discrete vectors of the received signal, transmit signal, and \ac{AWGN} signal; $\ell_p \triangleq \lfloor \frac{\tau_p}{T_\mathrm{s}} \rceil \in \mathbb{N}_0$ and $f_p \triangleq \frac{N\nu_p}{f_\mathrm{s}} \in \mathbb{R}$ are the normalized integer\footnote{It is assumed that the sampling frequency $f_\mathrm{s}$ is sufficiently high, such that the normalized path delays can be assumed to be integers with negligble error \cite{Hong_2022,Bemani_TWC23,Rou_arxiv24}.} path delay and normalized digital Doppler shift of the $p$-th propagation path; and $\mathbf{H} \triangleq \sum_{p=1}^{P}h_p \!\cdot\! \mathbf{\Phi}_{p} \!\cdot\! \mathbf{Z}^{f_p} \!\cdot\! \mathbf{\Pi}^{\ell_p} \in \mathbb{C}^{N \times N}$ is the circular convolutional matrix of the doubly-dispersive channel, whose $p$-th path is fully parametrized of four statistical components: 
\textit{a)} a complex channel fading coefficient $h_p \in \mathbb{C}$, 
\textit{b)} a diagonal phase matrix $\boldsymbol{\Phi}_p  \!\in\! \mathbb{C}^{N \times N}$ given by eq. \eqref{eq:CCP_phase_matrix} corresponding to the chirp-periodic phase of the \ac{AFDM} \ac{CP} \cite{Bemani_TWC23},
\textit{c)} a diagonal roots-of-unity matrix $\mathbf{Z} \!\in\! \mathbb{C}^{N \times N}$ given by eq. \eqref{eq:Z_matrix} which is taken to the $f_p$-th power,
and \textit{d)} a right-multiplying circular left-shift matrix $\mathbf{\Pi} \in \mathbb{C}^{N \times N}$ 
%
%
taken to the $\ell_p$-th integer power, to denote a circular left-shift operation by $\ell_p$ indices.

In light of the above, the effective input-output relationship of the \ac{AFDM} system is given by \vspace{-0.5ex}
\begin{equation}
\mathbf{y} \triangleq \mathbf{G}\mathbf{x} + \tilde{\mathbf{w}} = \mathbf{A}\mathbf{r} \in \mathbb{C}^{N \times 1},
\label{eq:AFDM_IO_system} \vspace{-0.5ex}
\end{equation}
where $\mathbf{y} \in \mathbb{C}^{N \times1}$ is the demodulated signal, $\mathbf{G} \triangleq \mathbf{A}\mathbf{H}\mathbf{A}^{-1} \in \mathbb{C}^{N \!\times\! N}$ is the effective \ac{AFDM} channel matrix, and $\tilde{\mathbf{w}} \triangleq \mathbf{A}\mathbf{w} \in \mathbb{C}^{N \times 1}$ is the effective noise which has the same statistical properties as $\mathbf{w}$ due to unitary transform property of the \ac{DAFT}.

The unique property of the \ac{AFDM} effective channel $\mathbf{G}$ is the distinct separability of the channel paths with different integer delay and Doppler shifts \cite{Bemani_TWC23,Rou_arxiv24}, leading to diversity optimality and double-dispersion robustness in \ac{LTV} channels \cite{Bemani_TWC23}, given that the \ac{DAFT} chirp frequencies in eq. \eqref{eq:DAFT_matrices} are selected to satisfy
\begin{equation}
c_1^\mathrm{opt} =\frac{2(f^\mathrm{max} + \xi) + 1}{2N} ~\text{ and }~ c_2^\mathrm{opt} \in \{\mathbb{R} \backslash  \mathbb{Q}\},
\label{eq:AFDM_chirp_condition}
\end{equation}
where $\xi \in \mathbb{N}_0$ is the guard width of the \ac{AFDM} which improves the robustness against fractional Doppler shifts \cite{Bemani_TWC23,Rou_arxiv24}, and $\{\mathbb{R} \backslash  \mathbb{Q}\}$ denotes the set of real irrational numbers.


\section{Proposed \ac{AFDM}-based \ac{CPIM}}
\label{sec:proposed_CPIM}

In light of eq. \eqref{eq:AFDM_chirp_condition}, it can be seen that while the first chirp matrix $\mathbf{\Lambda}_{c_1}$ of the \ac{DAFT} is of critical value in determining the performance over the doubly-dispersive channel, the second chirp matrix $\mathbf{\Lambda}_{c_2}$ provides a flexible parameter while still attaining the desired properties of the \ac{AFDM}.
This degree of freedom is unique to the \ac{AFDM} modulation scheme, as other waveforms which achieve full diversity and robustness over doubly-dispersive channels, such as the \ac{OTFS} waveform \cite{Wei_WC21}, do not have such flexibility in the core transform used in the modulation.

Therefore, we propose a novel \ac{IM} scheme based on the second chirp matrix $\mathbf{\Lambda}_{c_2}$ of the \ac{AFDM}, specifically by exploiting the permutations of the chirp sequence $\boldsymbol{\lambda}_{c_2}$.

\subsection{The \ac{AFDM}-\ac{CPIM} Transmitter}
\label{sec:AFDM-CPIM}

Let the permutation operation of an arbitrary vector $\mathbf{x} \in \mathbb{C}^{N \times 1}$ be denoted by $\mathbf{x}_{i} \triangleq \mathrm{perm}(\mathbf{x}, i)$, where $i \in \{1,\cdots\!,N!\}$ is the permutation \textit{index} in ascending order
.
Then, a modified \ac{DAFT} matrix is described by the $i$-th permutation the second chirp sequence, \textit{i.e.,} \vspace{-0.5ex}
\begin{equation}
{\mathbf{A}}_i \triangleq {\mathbf{\Lambda}}_{c_2,i} \!\cdot\! \mathbf{F}_N \!\cdot\! \mathbf{\Lambda}_{c_1} \in \mathbb{C}^{N \times N},
\label{eq:modified_DAFT} \vspace{-1ex}
\end{equation}
where $\mathbf{\Lambda}_{c_2,i} \triangleq \mathrm{diag}(\boldsymbol{\lambda}_{c_2, i})$, and ${\boldsymbol{\lambda}}_{c_2, i} \triangleq \mathrm{perm}(\boldsymbol{\lambda}_{c_2}, i)$.

Trivially, there exist $N!$ number of permuted \ac{DAFT} matrices, which can be used in place of the original \ac{DAFT} matrix $\mathbf{A} = \mathbf{A}_1$ of the \ac{AFDM} modulator, without changing any of the delay-Doppler and diversity properties of the \ac{AFDM} waveform.
Therefore, in the proposed \ac{AFDM}-\ac{CPIM} scheme, additional information is encoded by selecting a specific permuted \ac{DAFT} matrix to modulate the symbols from a given codebook of $K$ unique permuted \ac{DAFT} matrices.
Namely, the permuted \ac{DAFT} matrix codebook is described by $\mathcal{A} \triangleq  \{{\mathbf{A}}_{q_k}\}_{k = 1}^{K}$ whose selected index of $K$ permutations is described by the index vector $\mathbf{q} \triangleq \{q_1, \cdots, q_k,\cdots,q_K\}$, where $k^* \in \mathcal{K} \triangleq \{1,\cdots\!,K\}$ denotes the specifically selected index for transmission.
Clearly, $K$ can be any value between $K = 2$ (binary \ac{CPIM}) and $K = 2^{\log_2(N!)}$, which for even moderate values of $N$ becomes immensely large, highlighting the potential of the proposed method.

\begin{figure}[t]
\centering
\noindent
\includegraphics[width=1\columnwidth]{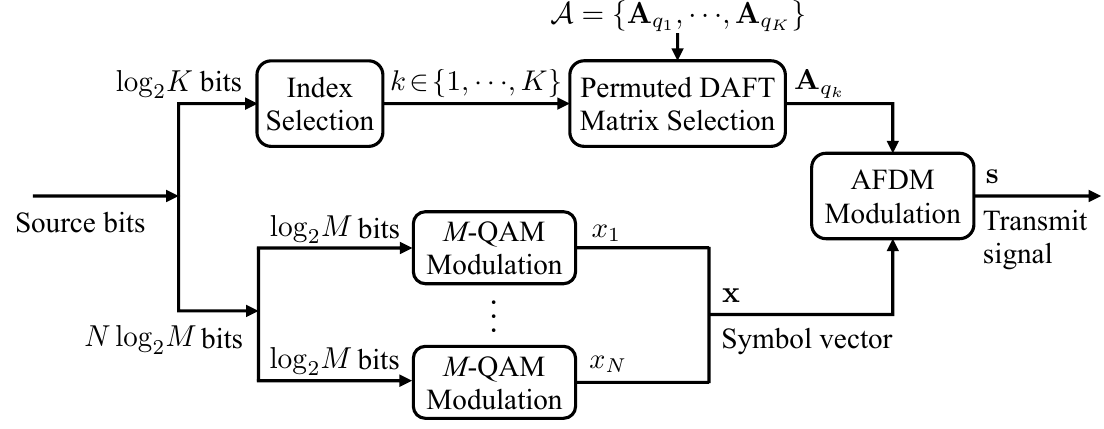}
\caption{A schematic diagram of the modulation process of the proposed \ac{AFDM}-\ac{CPIM} transmitter.}
\label{fig:schematic_CPIM}
\vspace{-2ex}
\end{figure}

In summary, the \ac{AFDM}-\ac{CPIM} efficiently encodes a source bit sequence of length $B \triangleq N\log_2{M} + \log_2(K)$ into the transmit signal by using the first $B_1 \triangleq N\log_2{M}$ bits to map $N$ complex symbols from an $M$-ary constellation, then using the remaining $B_2 \triangleq \log_2(K)$ bits to select the index $k^*$ of the permuted \ac{DAFT} matrix from the codebok $\mathcal{A}$ used to modulate the symbols, as illustrated in a schematic diagram in Fig. \ref{fig:schematic_CPIM}.

\subsection{MMSE-based Reduced-Complexity Detector}
\label{sec:MMSE-ML_detector}

The complementary \ac{AFDM}-\ac{CPIM} detector with perfect \ac{CSI} $\mathbf{H}$, consequently must detect the correct \ac{DAFT} matrix from a codebook of $K$ valid matrices, in addition to the $N$ complex symbols from an $M$-ary constellation $\mathcal{X}$.
The corresponding \ac{ML} problem is given by

\begin{align}
\tilde{\mathbf{x}}, \tilde{k}^* =\!  \underset{\substack{k \in \mathcal{K} \\ ~\mathbf{x} \in \mathcal{X}^{N}}}{\mathrm{argmin}} \big\| \mathbf{y}_k \!-\! \mathbf{A}_k^{\vphantom{-1}}\mathbf{H} \mathbf{A}^{\!-1}_k {\mathbf{x}} \big\|_2^2 \!=\! \underset{\substack{k \in \mathcal{K}  \\ ~\mathbf{x} \in \mathcal{X}^{N}}}{\mathrm{argmin}} \big\| \mathbf{r} \!-\! \mathbf{H} \mathbf{A}^{\!-1}_k {\mathbf{x}} \big\|_2^2, \nonumber \\[-3ex]
\label{eq:ML_metric}
\end{align} 

\vspace{-0.5ex}
\noindent where $\mathbf{y}_k \triangleq \mathbf{A}_k \mathbf{r} \in \mathbb{C}^{N \times 1}$ is the demodulated \ac{AFDM} signal with the $k$-th permutated \ac{DAFT} matrix.

However, the above \ac{ML} detection requires a discrete search space $K M^N$, such that the detection complexity order is $\mathcal{O}(KM^N)$, which is evident to quickly become infeasible for practical communications systems with large number of subcarriers $N$.
Therefore in the following, we propose a novel reduced-complexity \ac{ML} detector leveraging a matched \ac{MMSE} filter bank.

The \ac{MMSE} filter of the system for $\mathbf{x}$ in eq. \eqref{eq:AFDM_IO_system} with the $k$-th permuted \ac{DAFT} matrix $\mathbf{A}_k$, is given by 
\begin{align}
\mathbf{M}_k & \triangleq (\mathbf{H}\mathbf{A}^{\!-1}_k)\herm \big((\mathbf{H}\mathbf{A}^{\!-1}_k) (\mathbf{H}\mathbf{A}^{\!-1}_k)\herm + N_0 \mathbf{I}_N)\big)^{-1} \nonumber \\
& = \mathbf{A}_k \mathbf{H}\herm \big(\mathbf{H} \mathbf{H}\herm + N_0 \mathbf{I}_N)\big)^{-1} \in \mathbb{C}^{N \times N}
\label{eq:MMSE_filter}
\end{align}

\vspace{-0.5ex}
\noindent where $N_0$ is the spectral density of the \ac{AWGN} vector $\mathbf{w}$.

In turn, the estimated symbol vector with the $k$-th \ac{MMSE} filter is given by \vspace{-0.5ex}
\begin{equation}
\tilde{\mathbf{x}}^\mathrm{MMSE}_k \triangleq \mathcal{P}_{\mathcal{X}}(\mathbf{M}_k \mathbf{r}),
\label{eq:symbol_projection}
\end{equation}
where $\mathbf{M}_k \mathbf{r} \in \mathbb{C}^{N \times 1}$ is the raw \ac{MMSE}-equalized symbol vector, \textit{i.e.,} the soft-estimate of the symbol vector, and $\mathcal{P}_{\mathcal{X}}(\,\cdot\,)$ is the projection operator to the discrete set $\mathcal{X}$, \textit{i.e.,} $\tilde{\mathbf{x}}^\mathrm{MMSE}_k$ is the hard-decided symbol vector.

In light of the above, the optimal $k$-th permutation is determined by first applying all $K$ \ac{MMSE} filters to the received signal, then solving the reduced \ac{ML} problem as
\begin{equation}
\tilde{k}^{*} = \underset{k \in \mathcal{K} }{\mathrm{argmin}} \big\| \mathbf{r} - \mathbf{H}\mathbf{A}_k \tilde{\mathbf{x}}^\mathrm{MMSE}_i \big\|_2^2,
\label{eq:MMSE_ML_metric}
\vspace{-0.5ex}
\end{equation}
which only requires $K$ evaluations of the \ac{ML} metric, albeit the $K$ applications of the \ac{MMSE} filter such that the detection complexity is of order $\mathcal{O}(KN^3)$.

The proposed reduced-complexity \ac{MMSE}-\ac{ML} detector is fully described in Algorithm \ref{alg:MMSEML}, which can achieve a complexity order reduction from $\mathcal{O}(KM^N)$ of the full \ac{ML} problem in eq. \eqref{eq:ML_metric} to $\mathcal{O}(KN^3)$, with a trade-off in optimality of the solution in relaxing the \ac{ML} symbol estimation of via the \ac{MMSE} estimator.
Consequently, in Section \ref{sec:quantum_accelerated_ML_detection}, a quantum algorithm is designed to address the full \ac{ML} problem of eq. \eqref{eq:ML_metric} to alleviate the challenging computational complexity and enable the full potential of the proposed \ac{AFDM}-\ac{CPIM} scheme.

\begin{algorithm}[H]
\hrulefill
\vspace{-0.75ex}
\begin{algorithmic}[1]
\Statex \hspace{-3.25ex} {\bf{Inputs:}} Received signal $\mathbf{r}$, permuted \ac{DAFT} Codebook $\mathcal{A}$,  
\Statex \hspace{4.67ex} constellation $\!\mathcal{X}\!$, channel matrix $\!\mathbf{H}$, noise variance $\!N_0$.
\Statex \hspace{-3.5ex} {\bf{Outputs:}} Estimated symbol vector $\tilde{\mathbf{x}}$ and codeword index $\tilde{k}^*$
\vspace{-4ex}
\Statex \hspace{-4ex}\hrulefill
%
%
\State Compute the \ac{MMSE} filters $\mathbf{M}_k$ $\forall k$ via eq \eqref{eq:MMSE_filter}; 
\State Obtain the \ac{MMSE}-equalized vector $\tilde{\mathbf{x}}_k$ $\forall k$ via eq. \eqref{eq:symbol_projection};
\State Compute the \ac{ML} metrics $\forall k$ in eq. \eqref{eq:MMSE_ML_metric};
\State Determine the index $k_\mathrm{min}$ with the smallest \ac{ML} metric;
\State Output $\tilde{k}^* \leftarrow k_\mathrm{min}$ as the estimated codeword;
\State Output $\tilde{\mathbf{x}} \,\leftarrow \tilde{\mathbf{x}}^\mathrm{MMSE}_{k_\mathrm{min}}$ as the estimated symbol vector; 

\end{algorithmic}
\caption[]{\!\!: Proposed \ac{MMSE}-\ac{ML} \ac{AFDM}-\ac{CPIM} Detector}
\label{alg:MMSEML}
\end{algorithm}

\section{Quantum-Acceleration for \ac{AFDM}-\ac{CPIM}}
\label{sec:quantum_acceleration}

In this section, two quantum algorithms are formulated to address the optimal design problem of the permuted \ac{DAFT} matrix codebook of \ac{AFDM}-\ac{CPIM} systems, and the full \ac{ML} detection problem as outlined in eq. \eqref{eq:ML_metric}, which are highly challenging due to the inherently large discrete solution space.
Specifically, we reformulate the two problems as a polynomial binary optimization problem, which enable its solution to be obtained via the \ac{GAS} quantum algorithm, thus surpassing the computational complexity barrier of classical optimization.

In particular, \ac{GAS} \cite{gilliam2021grover} can solve binary optimization problems with a query complexity $\mathcal{O}(\sqrt{N})$, where $N = 2^n$ is the size of the search space and $n$ the number of binary variables.
The procedure of \ac{GAS} is described in Algorithm~\ref{alg:GAS} below, assuming a fault-tolerant quantum computer equipped with $n+m$ qubits, where $m$ qubits are used for encoding a general objective function value $E(\mathbf{b})$.
The method for constructing a quantum circuit of \ac{GAS} for polynomial binary optimization problems, consisting of a state preparation operator $\mathbf{S}_{y_i}$ and the Grover rotation $\mathbf{R}$, was introduced in \cite{gilliam2021grover}, extended to real-value variables in \cite{norimoto2023quantum} and tested on a real quantum computer.
%

\subsection{Quantum-Accelerated Optimized Codebook Design}
\label{sec:quantum_codebook_search}

In the proposed \ac{AFDM}-\ac{CPIM} scheme, $K$ permuted \ac{DAFT} matrices out of $N!$ candidates are selected as valid codewords, which can convey additional $\log_2(K)$-bit of information. 
Here, the system performance can be improved by carefully selecting the codewords according to an appropriate distance metric.

It is well known that the minimum distance between two distinct codewords $d_{\mathrm{min}}$, generally becomes the dominant performance factor of wireless communication systems, especially in middle to high SNR region.
Therefore, the problem of designing an optimized codebook can be defined as follows.

Given a distance $d_{i, j}$ between a pair of codewords ${\mathbf{A}}_i$ and ${\mathbf{A}}_j$, it is required to find a codebook $\mathcal{A}$ such that $|\mathcal{A}| = K$ and $d_{\mathrm{min}} = \min_{i \neq j} \qty{d_{i, j} \mid \qty{\mathbf{A}_i, \mathbf{A}_j} \subseteq \mathcal{A}}$ is maximized. 
This problem can be formulated as a polynomial binary optimization problem \cite{gilliam2021grover} with the objective function \vspace{-0.5ex}
\begin{equation}
E(\mathbf{b}) = \sum_{i, j}  b_i b_j\big((d_{i, j})^{-1}\big)^{\lambda_1} \!+ \lambda_2 \big(\sum_i b_i - K\big)^2,
\label{eq:objfun_codebook}
\vspace{-1.5ex}
\end{equation}

\begin{algorithm}[H]
\hrulefill
\vspace{-1ex}
\begin{algorithmic}[1]
\Statex \hspace{-3ex} {\bf{Inputs:}} Polynomial binary objective function $E:\mathbb{F}_2^n\rightarrow\mathbb{Z}$ \vspace{-0.25ex}
\Statex \hspace{4.67ex} w/ \# of binary variables $n$, scaling factor $\lambda = 8/7$. \vspace{-0.25ex}
\Statex \hspace{-3ex} {\bf{Outputs:}} Binary vector solution $\mathbf{b}$.
\vspace{-1.75ex}
\Statex \hspace{-4ex}\hrulefill \vspace{-0.25ex}
\State {Sample $\mathbf{b}_0 \in \mathbb{F}_{2}^n$, set $y_0 = E(\mathbf{b}_0)$, $k=1$, and $i=1$}; \vspace{-0.25ex}
\Repeat \vspace{-0.25ex}
\State Randomly select the rotation count $L_i$ from the set \vspace{-0.25ex}
\Statex \hspace{3ex} $\{0, 1, \cdots, \lceil k-1 \rceil\}$;\vspace{-0.25ex}
\State Evaluate $\mathbf{R}^{L_i} \mathbf{S}_{y_i} \ket{0}_{n + m}$ to obtain $\mathbf{b}$ and $y = E(\mathbf{b})$;  \vspace{-0.25ex}
\If{$y < y_i$} \vspace{-0.25ex}
\State $\mathbf{b}_{i+1} = \mathbf{b}$, \;\,$y_{i+1} = y$, \;and $k = 1$; \vspace{-0.5ex}
\Else \vspace{-0.5ex}
\State $\mathbf{b}_{i+1}  = \mathbf{b}_i$, \,$y_{i+1} = y_i$, and $k = \min\{\lambda k, \sqrt{2^n}\}$; \vspace{-0.15ex}
\EndIf \vspace{-0.2ex}
\State $i = i + 1$; \vspace{-0.25ex}
\Until{a termination condition is met} \vspace{-0.25ex}
\end{algorithmic}
\caption[]{\!\!: \ac{GAS} Algorithm \cite{gilliam2021grover,norimoto2023quantum}}
\label{alg:GAS}
\end{algorithm}

\noindent where $\mathbf{b} = (b_1, \cdots\!, b_{N!})$ are the binary variables representing whether the $i$-th \ac{DAFT} matrix codeword $\mathbf{A}_i$ is included in the codebook or not, and $\lambda_1$ and $\lambda_2$ are penalty coefficients. 

The first term in eq. \eqref{eq:objfun_codebook} corresponds to the objective of maximizing $d_{\mathrm{min}}$, while the second term corresponds to the constraint of the codebook size $K$.
Then, assuming $d_{i, j} > 1$ for arbitrary indices $(i, j)$, and $\lambda_1 >\!\!> 1$, the minimum distance $d_{\mathrm{min}}$ of the optimal codebook solution of eq. \eqref{eq:objfun_codebook} is maximized. 
This can be easily justified as follows.

Let $d_{\mathrm{min}, 1}$ be the maximum possible $d_{\mathrm{min}}$, and $d_{\mathrm{min}, 2}$ be the second maximum possible $d_{\mathrm{min}}$. 
By ignoring the second term in eq. \eqref{eq:objfun_codebook} without the loss of generality, the upper bound of eq. \eqref{eq:objfun_codebook} for the codebook with $d_{\mathrm{min}} = d_{\mathrm{min}, 1}$ is given by $E(\mathbf{b}) \leq \bar{E}(\mathbf{b}) = K \big( (d_{\mathrm{min}, 1})^{-1}\big)^{\lambda_1}$, while the lower bound of eq. \eqref{eq:objfun_codebook} for the codebook with $d_{\mathrm{min}} \leq d_{\mathrm{min}, 2}$ is given by $E(\mathbf{b}) \geq \underline{E}(\mathbf{b}) = \big( (d_{\mathrm{min}, 2})^{-1}\big)^{\lambda_1}$.

When $\lambda_1$ is sufficiently large, $\bar{E}(\mathbf{b})$ is always less than $\underline{E}(\mathbf{b})$, and the optimal codebook obtained from eq. \eqref{eq:objfun_codebook} with the minimum objective function value satisfies $d_{\mathrm{min}} = d_{\mathrm{min}, 1}$.

Finally, by applying the aforementioned \ac{GAS} in Alg.~\ref{alg:GAS} to the above formulation, the query complexity is expressed as $\mathcal{O}\big(\sqrt{2^{N!}}\big)$, while the query complexity of the classical exhaustive search is given by $\mathcal{O}\big(\binom{N!}{K}\big)$.

\subsection{Quantum-Accelerated ML Detection}
\label{sec:quantum_accelerated_ML_detection}

In this section, the full \ac{ML} detection problem of \ac{AFDM}-\ac{CPIM} described in eq. \eqref{eq:ML_metric} is reformulated as a binary optimization problem using the approach proposed in \cite{norimoto2023quantum}, and a novel method to utilize multiple quantum devices in parallel in order to solve the intractable \ac{ML} problem is proposed.

Following the \ac{AFDM}-\ac{CPIM} transmitter structure of Sec. \ref{sec:AFDM-CPIM}, the symbol vector $\mathbf{x} \in \mathbb{C}^{N \times 1}$ can be described as being mapped from a bit sequence $\mathbf{b} = [b_1, \cdots\!, b_{B_1}] \in \mathbb{B}^{B_1}$ of length $B_1 = N\log_2{M}$, with the mapping function given by \vspace{-0.5ex}
\begin{equation}
\mathbf{x} = g(\mathbf{b}) = g(b_1,\cdots,b_{B_1}), \vspace{-0.5ex}
\end{equation}
as specified in the \ac{5GNR} standard \cite{3gpp2018ts}.

To provide one example, the \ac{BPSK} mapping function $\mathbf{x}=g_{\mathrm{BPSK}}(\mathbf{b})$ is given by \vspace{-0.5ex}
\begin{equation}
x_t=\tfrac{1}{\sqrt{2}}[(1-2b_t)+j(1-2b_t)], \vspace{-0.5ex}
\end{equation}
for $t = \{0, 1, \cdots, N-1\}$.
%
%
%

Next, we assume that $K$ quantum devices can be used in parallel to solve the \ac{ML} detection problem of eq. \eqref{eq:ML_metric}, where the binary objective function for the $k$-th device is denoted by \vspace{-1.5ex}
\begin{equation}
E^{(k)}(\mathbf{b}) \triangleq \norm{\mathbf{r}-\mathbf{H}\mathbf{A}_k^{-1}g(\mathbf{b})}^2_2.
\label{eq:MLD_GAS} \vspace{-0.5ex}
\end{equation}

By leveraging \ac{GAS} at each $k$-th quantum device, the binary solution ${\mathbf{b}}^{(k)}$ that minimizes the objective function $E^{(k)}$ is obtained.
Then, the optimal index ${k}^*$ is evaluated via comparing the optimal value of the objective function across the $K$ quantum devices and determining the minimum, as \vspace{-0.5ex}
\begin{equation}
{k}^* = \underset{k \in \mathcal{K}}{\mathrm{argmin}}~E^{(k)}({\mathbf{b}}^{(k)}), \vspace{-0.5ex}
\end{equation}
from which the optimal solution of the binary vector is consequently determined as ${\mathbf{b}}^{(k^*)}$.

The \ac{ML} detection problem of eq. \eqref{eq:MLD_GAS} can be formulated as a quadratic function for the \ac{BPSK} or \ac{QPSK} case, but must be formulated as a higher-order function for higher-order modulations, which cannot be solved by mathematical programming solvers such as CPLEX\footnotemark or by quantum annealing.
By contrast, the \ac{GAS} supports binary objective functions of any order \cite{norimoto2023quantum} and achieves a query complexity of $\mathcal{O}(K \sqrt{M^N})$, which is significantly lower than the classical computational complexity of $\mathcal{O}(K M^N)$. 

%
%
%

\vspace{-1ex}
\section{Performance Assessment}
\label{sec:performance_analysis}
\vspace{-1.5ex}

\begin{figure}[b]
\vspace{-0.1ex}
\centering
\noindent 
\includegraphics[width=0.98\columnwidth]{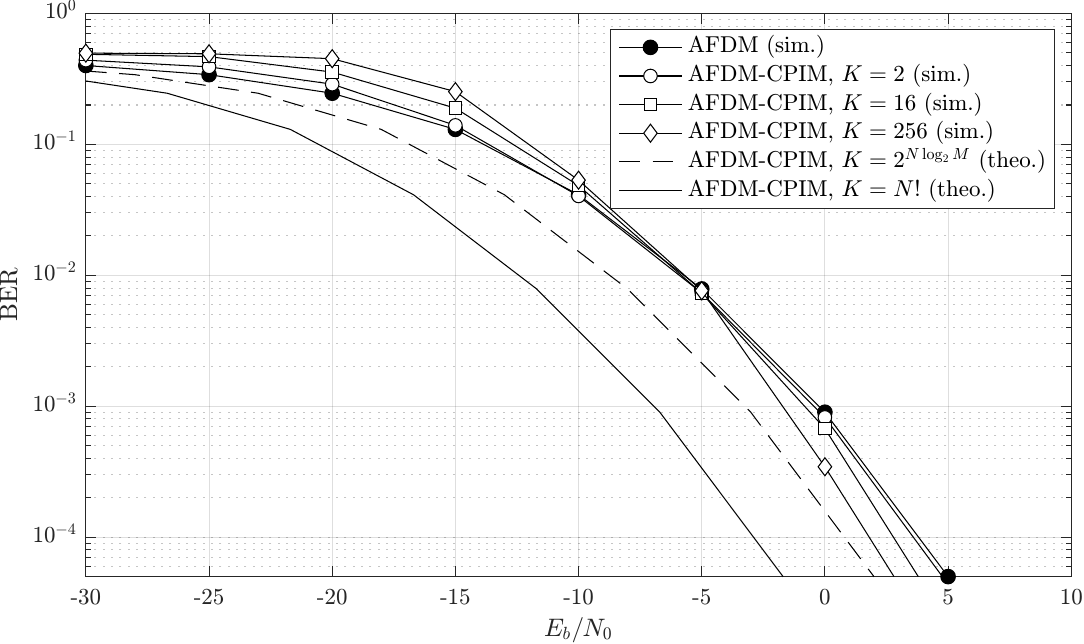}
\caption{\ac{BER} performance of \ac{AFDM}-\ac{CPIM} for different simulated and theoretical values of $K$, over a doubly-dispersive system with $P = 3, \ell^\mathrm{max} = 3,  f^\mathrm{max} = 3, N = 32, M = 2$.}
\label{fig:BER_perf}
\vspace{-2ex}
\end{figure}

In Fig. \ref{fig:BER_perf}, the performance of the proposed \ac{AFDM}-\ac{CPIM} is compared against the classical \ac{AFDM} in terms of the \ac{BER} with respect to the \ac{EbN0}, where the results have been obtained via the proposed \ac{MMSE}-\ac{ML} detector in Alg. \ref{alg:MMSEML}.

It is observed that the proposed \ac{AFDM}-\ac{CPIM} outperforms the classical \ac{AFDM} for sufficient \ac{EbN0}, benefiting from the additional chirp-permutation information bits embedded without the loss of subcarrier resources. 
Specifically, the theoretical \ac{EbN0} gain given $K$, can be obtained as $(1 + \frac{\log_2K}{N \log_2M}) \;\mathrm{dB}$, where $K$ can be maximally upto $N!$, yielding a maximum attainable gain of $6.7 \;\mathrm{dB}$ for the system in Fig. \ref{fig:BER_perf}.

However, a performance degredation is also visible for low \ac{EbN0}, owing to the imperfection of the \ac{MMSE}, although its effect is expected to be reduced for increasing system size $N$.

\begin{figure}[t]
\vspace{-2ex}
\centering
\subfloat[\footnotesize Codebook BER comparison \\ with $N = 8$, $K = 2,$ $M = 4$.]{%
\includegraphics[width=0.443\columnwidth,valign=b]{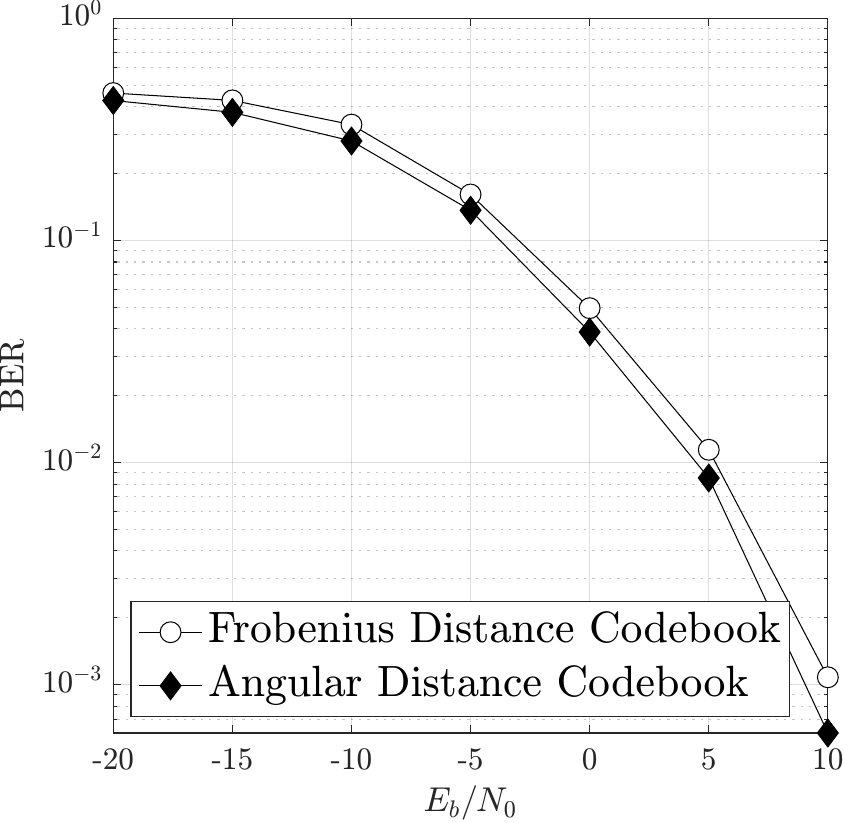}
\label{fig:codebook_BER}
}
\subfloat[\footnotesize Pairwise angular distances $d_{i,j}$, \\ with $N = 4$, $N!=24$, $K = 2$.]{%
\includegraphics[width=0.52\columnwidth,valign=b]{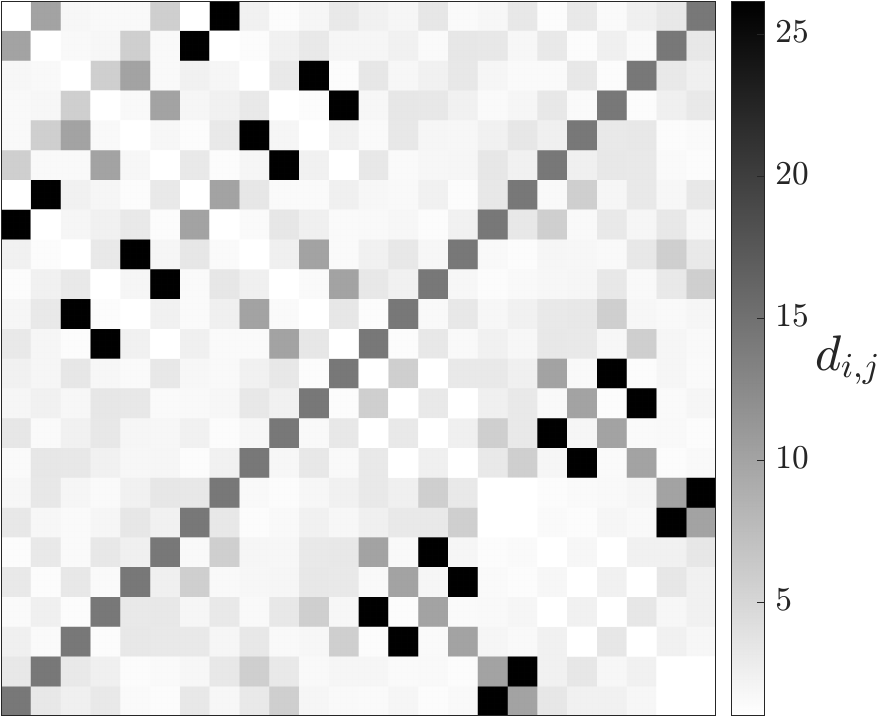}
\label{fig:codebook_matrix}
}
\vspace{-1ex}
\caption{\ac{BER} performance and pairwise distance analysis of the optimized \ac{AFDM}-\ac{CPIM} codebook.}
\label{fig:codebook}
\vspace{-3.5ex}
\end{figure}

Next, in Fig. \ref{fig:codebook_BER}, the \ac{BER} comparsion of two optimized codebooks under different exemplary distance metrics is provided, namely the Frobenius distance $d_{i,j} \triangleq ||\mathbf{A}_i-\mathbf{A}_j||_F$ and the angular distance $d_{i,j} \triangleq |{\tr(\mathbf{A}_i\herm \mathbf{A}_j)}|^{-1}$, where it is found that the latter provides a superior performance under the \ac{MMSE}-\ac{ML} detector.
However, the optimal codebook design should also consider the effective distance of the effective channels $\mathbf{G}_i$ and $\mathbf{G}_j$, and more elaborate distance metrics.

In addition, Fig. \ref{fig:codebook_matrix} presents the pairwise angular distances of a small system ($N = 4$) employing binary \ac{AFDM}-\ac{CPIM}, \textit{i.e.,} $K = 2$, which clearly illustrates certain codeword pairs with maximal distance.
The figure helps to anticipate the potential scale of the optimization problem for large $K$ and $N$, requiring a search of $K$ elements over the $N! \times N!$ grid.

\vspace{-4ex}
\section{Conclusion}

We proposed a novel \ac{IM} scheme based on a unique feature inherent to \ac{AFDM}.
The proposed \ac{AFDM}-\ac{CPIM} scheme is shown to provide a noticeable performance gain over conventional \ac{AFDM}, except for the computational complexity at large scales.
To address this issue, quantum-accelerated solutions are proposed both for the codebook design and \ac{ML} detection.
An analysis of the pairwise codeword distances suggests that the proposed \ac{CPIM} scheme can be exploited to realize efficient multi-user systems.
More details on the methods here proposed will be provided in the camera-ready, and the extension to multi-user scenarios will be exploited in a follow-up article.

\vspace{-1ex}

\footnotetext{\url{https://www.ibm.com/analytics/cplex-optimizer}} 

\end{document}